\documentclass[global,twocolumn]{svjour}
\usepackage{textcomp}
\usepackage{graphics}
\usepackage{epstopdf}
\usepackage{mathtools}
\usepackage{subfigure}
\usepackage{xcolor}
\usepackage{multirow}
\usepackage{cite}
\usepackage{ulem}
\usepackage{selinput}
\usepackage[english]{babel}
\usepackage{upgreek}
\usepackage{hyperref}
\usepackage{bm}
\journalname{}
\begin{document}
\title{Minimizing rf-induced excess micromotion of a trapped ion with the help of ultracold atoms}
\author{Amir Mohammadi, Joschka Wolf, Artjom Kr\"ukow, Markus Dei{\ss}, and Johannes Hecker Denschlag}

\institute{Institut f\"{u}r Quantenmaterie and Center for Integrated Quantum Science and
	Technology (IQ$^{ST}$),~Universit\"{a}t Ulm,~89069~Ulm, Germany\\Corresponding author: johannes.denschlag@uni-ulm.de}
\date{\today}
\authorrunning{A. Mohammadi et al.}
\titlerunning{Minimizing rf-induced excess micromotion of a trapped ion with the help of ultracold atoms}
\maketitle
\begin{abstract}
We report on the compensation of excess micromotion due to parasitic rf-electric fields in a Paul trap. The parasitic rf-electric fields stem from the Paul trap drive but cause excess micromotion, e.g. due to imperfections in the setup of the Paul trap. We compensate these fields by applying rf-voltages of the same frequency but adequate phases and amplitudes to Paul trap electrodes. The magnitude of micromotion is probed by studying elastic collision rates of the trapped ion with a gas of ultracold neutral atoms. Furthermore, we demonstrate that also reactive collisions can be used to quantify micromotion. We achieve compensation efficiencies of about 1$\:\text{Vm}^{-1}$, which is comparable to other conventional methods.
\end{abstract}

\section{Introduction}

Ideally, a single ion located in the center of a Paul trap experiences vanishing rf-trap fields, leading to vanishing micromotion. Typically, however, electrical stray fields and imperfections of the trap setup lead to a remaining level of micromotion, the excess micromotion. Minimization of this micromotion is important for many research fields such as quantum information processing \cite{Haeffner2008,Bermudez2017}, quantum simulation \cite{Johanning2009}, high precision spectroscopy \cite{Wineland1987,Wolf2008}, single-ion atomic clocks \cite{Ludlow2015}, and cold atom-ion collisions where reaching the $s$-wave regime is a challenge \cite{Tomza2017, Haerter2014, Zhang2017}. Therefore, in recent years much effort has been put into the investigation and minimization of micromotion. A variety of detection and compensation methods have been developed, which generally rely on optical probing the motional state of the ion (see, e.g. \cite{Berkeland1998, Allcock2010, Chuah2013, Gloger2015, Tanaka2012, Narayanan2011, Keller2015, Meir2017, Huber2014}). Recently, our group has demonstrated that excess micromotion due to static stray electrical fields can be sensitively probed and compensated with the help of a cold cloud of atoms which elastically collide with the ion \cite{Harter2013}. This method can also be applied to ions that are not laser-cooled. Furthermore, it is direction independent, in contrast to sideband techniques  as described e.g. in \cite{Berkeland1998}.

Here, we extend our work of \cite{Harter2013} and demonstrate the minimization of excess micromotion which is linked to rf-electric fields of the Paul trap. In particular, we compensate phase micromotion which is due to a time delay in the oscillating rf-voltages of opposite Paul trap electrodes. Furthermore, we compensate rf-induced micromotion along the axial direction of our linear Paul trap which can arise from rf-pick up on the endcap dc-electrodes or simply from imperfections in the alignment of electrodes. As a further development of the minimization method as compared to \cite{Harter2013} we show that instead of elastic collisions also reactive collisions between the ion and atoms can be used to probe excess micromotion. In fact, making use of the known scaling law of the reaction rate with collisional energy we can determine by which factor the kinetic energy of the ion is decreased. After minimization of the excess micromotion due to both dc- and rf-fields we estimate the remaining total residual excess rf-field amplitudes to be about 1$\:\text{Vm}^{-1}$ and the excess dc-fields to be about $0.02\:\text{Vm}^{-1}$. These compensation results are comparable to values reported using other methods \cite{Chuah2013, Tanaka2012, Narayanan2011, Keller2015, Meir2016}.

This article is structured as follows. In section \ref{sec:ion trap micromotion} we describe our ion trap setup and provide a brief review on excess micromotion. Then, in section \ref{sec:general_method} our detection method for micromotion is introduced. Section \ref{sec:elastic} is dedicated to the discussion of the minimization of phase and axial rf-excess micromotion, respectively. Here, atom loss due to elastic collisions is used as signal for optimization. In section \ref{sec: inelastic} we describe the probing of micromotion via reactive collisions. Finally, in section \ref{sec:summary} a summary is given and future prospects are addressed.

\section{Ion trap and excess micromotion}
\label{sec:ion trap micromotion}
\begin{figure}[t!]
	\begin{center}
		\includegraphics[width=\columnwidth]{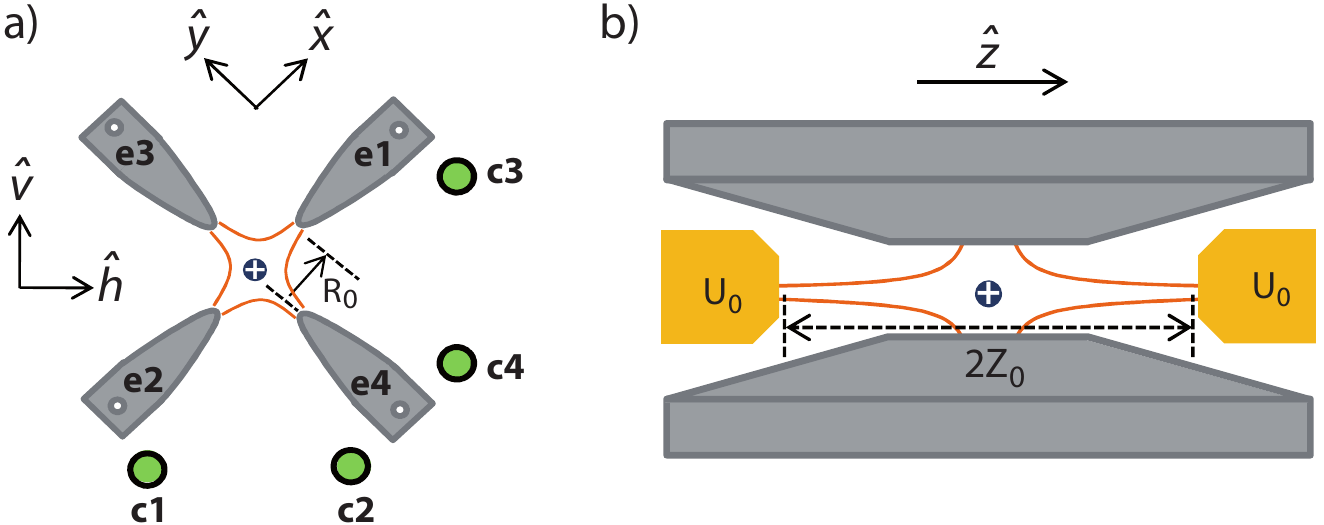}
		\caption{Schematic of the linear Paul trap. Shown is the configuration in the radial $\hat{x}$-$\hat{y}$-plane (\textbf{a}) and in the axial direction $\hat{z}$ (\textbf{b}). The rf-electrodes are indicated in gray, the endcap electrodes are yellow and the compensation electrodes (c1, c2, c3, c4) are green. For better visibility, the endcap electrodes are not depicted in \textbf{a} and the compensation electrodes are omitted in \textbf{b}. The orange lines illustrate some rf-electric field lines between the electrodes.}
		\label{fig:Paultrap}
	\end{center}
\end{figure}

In the following we consider micromotion in a linear Paul trap. Figure \ref{fig:Paultrap} shows the setup in our lab, which has been described in detail in \cite{Schmid2012, Haerter2013}. The four gray electrodes (e1-e4) are the rf-electrodes of the Paul trap. They are driven with a rf-frequency of $\Omega=2\pi \times 4.2\:\text{MHz}$ and generate the radial trapping confinement (i.e. within the $\hat{x}$-$\hat{y}$-plane) while the static field of two endcap electrodes (yellow) confines the ion in the axial direction ($\hat{z}$-direction). The effective distance from the trap center to the tips of the four rf-electrodes is $R_0 = 2.6\:\text{mm}$, while the spacing between the two endcap electrodes is $2 \times Z_0 = 14\:\text{mm}$. We nominally operate the ion trap in a symmetric manner\footnote{In a perfectly aligned linear Paul trap with vanishing rf-potentials on the endcaps such a symmetric rf-drive leads to a vanishing axial micromotion on the trap axis.
We note that in our previous work \cite{Haerter2013} we used an asymmetric trap drive.} where the rf-electrode pair (e1, e2) is driven by a voltage $+V_0 \, \cos(\Omega t)$ and the rf-electrode pair (e3, e4) by a voltage  $-V_0 \, \text{cos}(\Omega t)$. The voltage amplitude\footnote{Measurements of the electrode voltages indicate, however, that the voltage amplitude $V_0$ for the  electrode pair (e1, e2) and the pair (e3, e4) are not equal but  $160\:\text{V}$ and $143\:\text{V}$, respectively. Therefore, a cancellation of micromotion along the axial direction might be compromised. A simulation of the electrical fields for our setup can be found in \cite{Brunner2012}.}
is $V_0 \approx 150\:\text{V}$. At the center of the Paul trap this gives rise to the following electrical field
\begin{equation}
\label{eq:field1}
\begin{split}
\vec{\mathcal{E}}(x,y,z,t)=&-\frac{2V_0}{R'^2}(x\hat{x}-y\hat{y})\text{cos}(\Omega t)\\
&-\frac{\kappa U_0}{Z_0^2}(2z\hat{z}-x\hat{x}-y\hat{y})\,.
\end{split}
\end{equation}
We use $\{\hat{x},\hat{y},\hat{z}\}$ to denote the unit vectors for the directions of the coordinate system given in Fig.$\:$\ref{fig:Paultrap}. The first term in Eq.$\:$(\ref{eq:field1}) represents the electrical field generated by the rf-electrodes. Here, $R' \cong R_0$.  The second term expresses the electrical field due to the endcap electrodes, which are held at constant electrostatic potential $U_0 = 7.6\:\text{V}$, and $\kappa = 0.29 $ is a geometrical factor for our setup. We work with single $^{138}$Ba$^+$ ions at trapping frequencies (for the secular motion) of $\omega_{x,y,z}=2\pi\times (131,130,38.8)\:\text{kHz}$. The trap depth is about 2$\:$eV. Before each measurement the ion is laser-cooled to the Doppler limit.

For the ideal case of Eq.$\:$(\ref{eq:field1}), the particle is trapped exactly at the origin ($x = y = z = 0$) and exhibits vanishing micromotion. An additional quasi-constant  stray electrical offset field with components in the transverse direction, however, shifts the ion to a different position where it experiences oscillating electrical fields and therefore undergoes micromotion. Applying the experimental technique that we have demonstrated in Ref.$\:$\cite{Harter2013} such an electrical stray field can be very well compensated to values smaller than $0.02\:\text{Vm}^{-1}$ in our setup. This is the starting point for the measurements described in the present work. The kinetic energy contribution due to dc-stray electric fields is negligible as compared to the excess micromotion energies discussed in the remainder. In the following we simply assume the dc-stray fields to be fully compensated and we focus on excess micromotion resulting from phase delay and rf-fields in axial direction. A more general discussion can be found in the \hyperref[sec:Appendix]{Appendix}.

Excess micromotion due to phase delay occurs when there is a relative phase $\varphi_x$ between the oscillating voltages of the electrode pair (e1, e2) or a relative phase $\varphi_y$ for the pair (e3, e4). Using the same approach as in Ref.$\:$\cite{Berkeland1998}, such phase differences give rise to additional oscillating electrical field terms. For $\varphi_{x_i} \ll 1$ ($x_i \in \{x,y\}$) these terms can be approximated by
\begin{equation}
\label{eq:phase}
\vec{\mathcal{E}}_{x_i}= V_0 \frac{\alpha_{x_i}\varphi_{x_i}}{2R_0} \sin(\Omega t)\hat{x_i}\equiv \mathcal{E}_{x_i,0} \, \sin(\Omega t)\hat{x_i}\,,
\end{equation}
where the factors $\alpha_{x_i}$ depend on the trap geometry. We note that our configuration is characterized by $\alpha_{x_i}\approx 0.8$.

Rf-induced axial micromotion occurs when rf-electric fields are created along the trap axis, e.g. due to slight misalignment of the rf-electrodes or as a consequence of unwanted, asymmetric pick up of the rf-drive voltage on the endcap electrodes. This produces dominantly the electric field
\begin{equation}
\label{eq:acmicro}
\vec{\mathcal{E}}_z= {\mathcal{E}}_{z,0} \, \sin(\Omega t +\varphi_z)\ \hat z\,,
\end{equation}
in the trap center, with unknown amplitude ${\mathcal{E}}_{z,0}$ and phase $\varphi_z$.

Excess micromotion  readily increases the kinetic energy of the ion. This fact is also exploited for our detection scheme. The average kinetic energy is given by $E^\text{kin}=m_\text{Ba} \langle \dot{u}^2 \rangle/2$ (see also \cite{Berkeland1998}), where $m_\text{Ba}$ and $\dot{u}$ are the mass and the velocity of the Ba$^+$ ion, and $\langle \, \rangle$ represents the time average over a period of the secular motion at frequency $\Omega$. The individual components of $\dot{\bm{u}}$ are derived from the equations of motion. Using this approach, the kinetic energy contributions of motion in the three directions of space are given by
\begin{equation}
\label{eq:muEnergy}
E^\text{kin}_{x_i}=\frac{e^2\mathcal{E}_{x_i}^2}{4m_\text{Ba}\Omega^2}\,,
\end{equation}
where $e$ is the elementary charge and $x_i \in \{ x, y, z\}$.

The four green electrodes (c1-c4) in Fig.$\:$\ref{fig:Paultrap}a are used to compensate the ac-electric field components due to phase delay. We denote the pairs of compensation electrodes (c1, c2) and (c3, c4) as vertical ($\text{v}$) and horizontal ($\text{h}$) electrode pairs, respectively. Driving the vertical pair of compensation electrodes with the same ac-voltage will create an rf-electrical field at the position of the ion, which is pointing along the $\hat{\text{v}}$-direction. An ac-voltage applied at the horizontal pair of compensation electrodes produces an rf-electric field along $\hat{\text{h}}$. One of the endcaps is utilized to compensate ac-electric fields along the $z$-axis\footnote{We note that besides applying ac-voltages to compensation electrodes, also alternative compensation methods could be used. For example, phase delays can be implemented by adjusting cable lengths or by using additional capacitances similarly as in \cite{Herskind2009}.}.

Besides excess micromotion arising from dc-electrical stray fields and parasitic rf-fields, there is yet another kind of excess micromotion present in our setup. It is linked to elastic collisions of the ion with the cold atoms and has been predicted and investigated in \cite{Cetina2012, Tomza2017, Fuerst2018, Meir2016}. In simple terms its origin can be understood as follows: In a collision the ion can be pulled out from the center of the ideal Paul trap where no micromotion occurs to a location with non-vanishing electrical rf-fields and micromotion. Thus, even at negligible temperatures of the atom cloud the ion can acquire a non-vanishing average kinetic energy. The typical kinetic energy scale for collisional excess micromotion can be calculated. Using the approach of \cite{Cetina2012} for a 3D trap and taking our current trap parameters we obtain about 40$\:\upmu \text{K}\times k_\text{B}$ for a Ba$^+$ ion colliding with ultracold Rb atoms. Such collisional micromotion cannot be compensated. However, as it is a function of the atom-ion mass ratio and the general ion trap parameters, setups and configurations can be optimized to minimize it.

\section{General method for minimizing excess micromotion using cold atoms}
\label{sec:general_method}

In order to minimize excess micromotion we generalize here the method we introduced
in \cite{Harter2013}. A single, cold Ba$^+$ ion in a Paul trap is immersed into an ultracold cloud of Rb atoms. In the cloud elastic and reactive atom-ion collisions take place with rates that depend on the micromotion energy. We find the minimum of micromotion by steering towards a local minimum (maximum) for the elastic (reactive) rate, respectively. Tuning of the micromotion is done via suitable electrical rf-fields at the center of the Paul trap which are produced by applying rf-voltages on the compensation electrodes or on one of the endcap electrodes of the Paul trap.

The ultracold Rb atoms are held in a far-off-resonant, crossed optical-dipole trap at a wavelength of $1064\:\text{nm}$ with a trap depth of about $22\:\upmu\text{K}\times k_\text{B}$. The atomic temperature is about $700\:\text{nK}$. The atoms are spin-polarized in the hyperfine state $f=1,m_f=-1$. We work with $1$ to $3\times 10^4$ $^{87}$Rb atoms and shot-to-shot fluctuations of the atom number are typically on the level of a few percent.

Once it is immersed into an atomic cloud, a trapped ion undergoes elastic collisions with the atoms, which quickly leads to a non-thermal kinetic energy distribution of the ion. Because of its relevance for the developing field of cold atom-ion interactions \cite{Tomza2017, Haerter2014, Zhang2017}, this issue has been recently investigated in a number of studies (e.g. \cite{Nguyen2012, Hoeltkemeier2016, Zipkes2011, Meir2016, Haze2018, Ravi2012, Dutta2017, Rouse2018}). The ionic energy distribution depends in a non-trivial way on quantities such as the atom-ion mass ratio, the atomic cloud size, and the ion trap parameters. In our case of Rb and Ba$^+$, the kinetic energy distribution of the ion is still nearly thermal, and the ion's average kinetic energy $E^\text{kin,a}$ in the presence of atoms is given approximately by $5\times E^\text{kin}$, where $E^\text{kin}$ represents the excess micromotion energy in the absence of atoms \cite{Krukow2016a}. When we only compensate excess micromotion due to dc-electrical fields the remaining kinetic energy of the ion is about $E^\text{kin,a}=4\:\text{mK}\times k_\text{B}$ \cite{Krukow2016a,Krukow2016b} in our trap. This energy is partially due to phase delay and axial rf-fields.

\section{Probing micromotion compensation via elastic atom-ion collisions}
\label{sec:elastic}

Here, we describe how we use elastic collisions between atoms and the ion to minimize rf-induced excess micromotion. As already discussed the typical kinetic energy of the ion in the atomic cloud is in the range of a few $\text{mK}\times k_\text{B}$ for our experiments. Thus, when the ion elastically collides with an ultracold atom it will typically kick the atom out of its shallow dipole trap. Alternatively, it only heats the atomic cloud at first, which finally also leads to atomic loss due to evaporation. In general, we expect the atomic loss to increase weakly with the average ion energy and therefore with the excess micromotion. The rate $\Gamma_\text{el}$ for elastic binary collisions of an atom and an ion with reduced mass $\mu$ is given by \cite{Cote2000}
\begin{equation}
\label{eq:elasticrate}
\Gamma_\text{el}=\sigma_\text{el}n_\text{at}\sqrt{2E_\text{col}/\mu} \propto E_\text{col}^{1/6}\,,
\end{equation}
where $n_\text{at}$ is the atomic particle density, $\sigma_\text{el}$ is the atom-ion elastic scattering cross section and $E_\text{col}$ is the two-body collision energy of the particles in the center-of-mass reference frame. Furthermore, we have used
\begin{equation}
\label{eq:elastic}
\sigma_\text{el} = \pi \left( { \mu C_4^2 \over \hbar^2 } \right)^{1/3} \left(1 + { \pi^2 \over 16 } \right) E_\text{col}^{-1/3}\,,
\end{equation}
as derived from a semiclassical approach \cite{Cote2000}. The constant
$C_4 = \alpha_{\text{Rb}} e^2/({4\pi \varepsilon_0})$ is defined via the polarization potential between the Rb atom and the ion, $V_\text{pol} = -C_4/(2 r^4)$. Here, $\alpha_{\text{Rb}} =  47.39(8)$\AA$^3$ \cite{Greg2015}, $r$ is the distance between atom and ion, and $\varepsilon_0$ is the vacuum permittivity. In order to probe excess micromotion, we measure the loss rate of the atom number in the atomic cloud for various rf-voltages applied on the respective electrodes used for compensation.

Besides the elastic collisions between atom and ion also inelastic and reactive collisions can take place which disturb our minimization scheme. A typical reactive process is the three-body recombination of $\text{Ba}^++\text{Rb}+\text{Rb}$ for which the reaction rate is given by $\Gamma_\text{inel}=k_3n_\text{at}^2$ with $k_3=1.04\times 10^{-24}\:\text{cm}^6\text{s}^{-1}$ for a three-body collisional energy of $2.2\:\text{mK}\times k_\text{B}$ \cite{Krukow2016b}. Another reaction is charge exchange, $\text{Ba}^++\text{Rb}\rightarrow \text{Ba}+\text{Rb}^+$, with an energy independent rate $k_2 n_\text{at}$, where $k_2=3.1\times 10^{-13}\:\text{cm}^3\text{s}^{-1}$ \cite{Krukow2016b}. In order to suppress three-body recombination we work with comparatively low densities $n_\text{at}$ ranging from $2$ to $4 \times 10^{11}\:\text{cm}^{-3}$. This reduces the total reaction rate to about $0.3\:\text{Hz}$. Nevertheless, since for our experiments typical interaction times of up to 1$\:$s are needed in order to gain enough atom loss due to elastic collisions, there is still a sizeable probability that the Ba$^+$ ion undergoes a reaction. We therefore use post-selection to only take into account runs where no reaction between the Ba$^+$ ion and an atom has occurred. For this, we determine via fluorescence imaging whether the Ba$^+$ ion is still present in the trap center immediately after the interaction time with the atom cloud. All runs for which this is not the case, are discarded.

\subsection{Compensation of phase micromotion}
\label{sec: phase}

Following Eq.$\:$(\ref{eq:phase}) we compensate transverse phase micromotion by applying suitable voltages $V_\text{c,h} \sin(\Omega t)$ and $V_\text{c,v} \sin(\Omega t)$ to the compensation electrode pairs $\text{h}$ and $\text{v}$. These rf-voltages are added on top of the dc compensation voltages. For this, we use a two-channel signal generator which is phase-locked to the rf-drive of the Paul trap.

\begin{figure}[t]
	\includegraphics[width=\columnwidth]{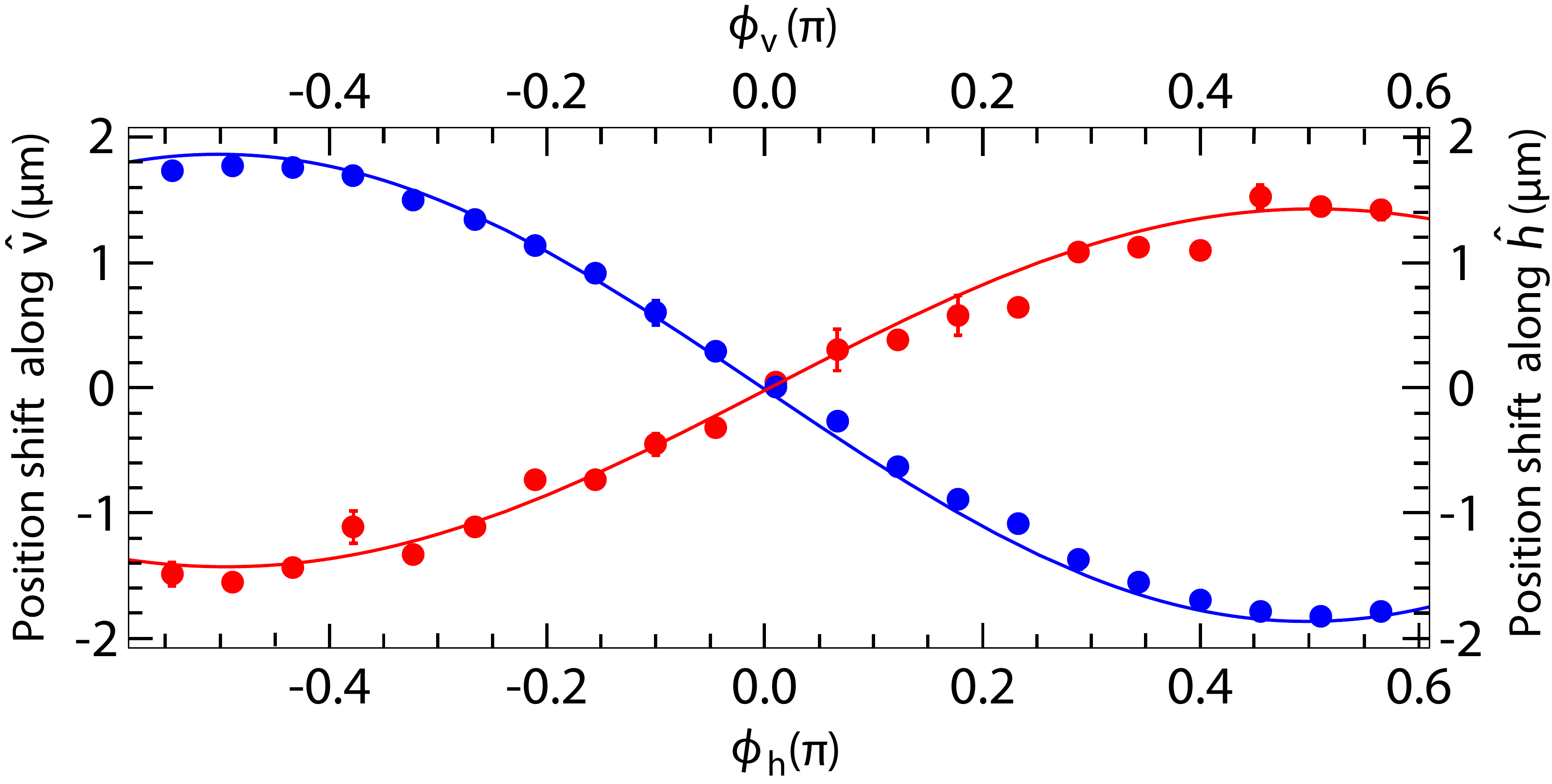}
	\caption{Measured position shifts of a Ba$^+$ ion along the directions $\hat{\text{v}}$ (blue filled circles) and $\hat{\text{h}}$ (red filled circles) (see Fig.$\:$\ref{fig:Paultrap}a) as functions of the phases $\phi_\text{h}$ and $\phi_\text{v}$, respectively. The solid lines are functions $\propto-\text{sin}(\phi_\text{h})$ (blue) and $\propto \text{sin}(\phi_\text{v})$ (red).}
	\label{fig:phase}
\end{figure}

Because we do not precisely know the phases of the rf-compensation voltages at the location of the respective electrodes we first carry out calibration measurements to determine these phases. We use the fact, that according to the trigonometric addition formulas a phase deviation of the compensation voltage, i.e. $ \propto \sin(\Omega t + \phi)$ can be written as $\propto [\sin(\phi) \cos(\Omega t)+\cos(\phi) \sin(\Omega t)]$. The component $\propto \sin(\phi) \cos(\Omega t)$ leads to a position shift [see Eq.$\:$(\ref{eq:field1})] and the corresponding spatial displacement of the ion is proportional to $\sin(\phi)$. Instead of only a single phase $\phi$, there are two different phases $\phi_\text{h}$ and $\phi_\text{v}$ in our experiment, associated with the compensation electrode pairs $\text{h}$ and $\text{v}$. A phase $\phi_\text{h}$ ($\phi_\text{v}$) leads to a displacement in the $\hat{\text{v}}$ ($\hat{\text{h}}$) -direction, respectively. We carry out two measurements of the Ba$^+$ ion position (using fluorescence imaging) where we vary either $\phi_\text{h}$ or $\phi_\text{v}$.  Figure \ref{fig:phase} shows the data. For simplicity, we have defined the phases $\phi_\text{h}$ and $\phi_\text{v}$ such that $\phi_\text{v}=0$ and $\phi_\text{h}=0$ correspond to a vanishing position shift. Both data sets were obtained using ac-voltage amplitudes of $V_\text{c,v} = 10\:\text{V}$ ($V_\text{c,h}=10\:\text{V}$), respectively, on the compensation rods, individually creating an electric field amplitude of $31\:\text{Vm}^{-1}$ at the trap center. In order to obtain larger position shifts of the ion, the voltage amplitude for the quadrupole blades was reduced by about a factor of 0.6. With these parameters the maximum position shifts are between $1$ to $2\:\upmu\text{m}$. We note that the line of view of the camera detecting the ion is perpendicular to the $\hat{\text{v}}$-direction but has an angle of about $45^\circ$ with respect to the $\hat{\text{h}}$-direction. Therefore, the position shift along the $\hat{\text{h}}$-axis appears smaller than it is.

\begin{figure}[t]
	\begin{center}
		\includegraphics[width=\columnwidth]{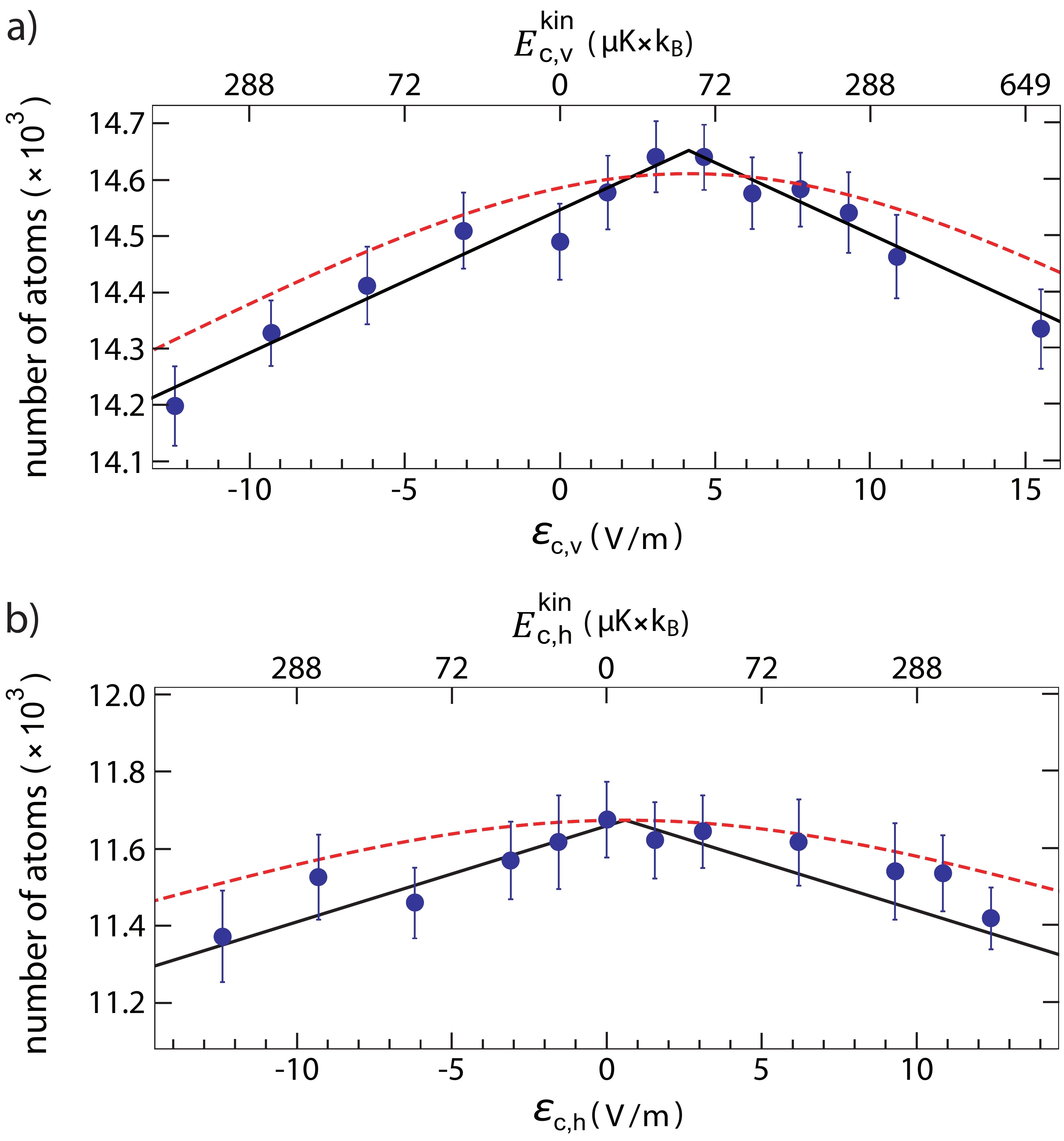}
		\caption{Remaining atom number after 1$\:$s of interaction with the ion as a function of the electric field amplitude $\varepsilon_\text{c,v}$ (\textbf{a}) and $\varepsilon_\text{c,h}$ (\textbf{b}), respectively. The upper horizontal axis translates the electric field amplitudes into the corresponding kinetic micromotion energies $E^\text{kin}_\text{c,h(v)}$ as determined by Eq.$\:$(\ref{eq:muEnergy}). Measurements are given by blue data points. Each of these data points is the average value of 170 experimental runs. The error bars represent the $1\sigma$ statistical uncertainty. Black solid lines are fits of the cusp function $N_j=-\chi_j |\varepsilon_\text{c,j}- \varepsilon_\text{c,j}^\text{max}|+N^\text{max}_j$ with $j\in \{\text{v,h}\}$. The red dashed lines represent model calculations for the remaining number of atoms for the given experimental parameters and assuming elastic two-body atom-ion collisions [Eq.$\:$(\ref{eq:remaining})]. In the model, also the background atom loss is taken into account (see text).
		}
		\label{fig:phasemini}
	\end{center}
\end{figure}

We now work with phases $\phi_\text{h}=0$ and $\phi_\text{v}=0$, and minimize phase micromotion. For this, we step through a range of ac-voltage amplitudes $V_\text{c,v}$ and $V_\text{c,h}$ and search for a minimum in atomic loss. Figures \ref{fig:phasemini}a and b show the remaining atom numbers as a function of the field amplitudes $\varepsilon_\text{c,v} =   V_\text{c,v} \times 3.1\:\text{m}^{-1}$ and $\varepsilon_\text{c,h} =  V_\text{c,h} \times 3.1\:\text{m}^{-1}$, respectively. For plot (a) we set $V_\text{c,h}=0$ while in (b) $V_\text{c,v}=0$ was used.
The interaction time was $1\:\text{s}$. We change the sign of the electric field by flipping its phase by $\pi$ at the rf-generator. On the top axis of the figures the electrical field amplitudes $\varepsilon_\text{c,h(v)}$ are translated into the corresponding micromotion energies $E^\text{kin}_\text{c,h(v)}$ using Eq.$\:($\ref{eq:muEnergy}).

As can be read off from Fig.$\:$\ref{fig:phasemini}a, an electric compensation field amplitude of  $\varepsilon_\text{c,v} \approx 4\:\text{Vm}^{-1}$ reduces the micromotion energy by about  $50\:\upmu\text{K}\times k_\text{B}$. In contrast to that, Fig.$\:$\ref{fig:phasemini}b reveals that micromotion in the horizontal direction is already close to the minimum such that only small compensation fields $\varepsilon_\text{c,h}$ are needed. In order to determine optimal electric field amplitudes for compensation more precisely, we heuristically use a cusp-like fit function
$N_j=-\chi_j |\varepsilon_\text{c,j}- \varepsilon_\text{c,j}^\text{max}|+N^\text{max}_j$ with $j\in \{\text{v,h}\}$ for the remaining atom numbers $N_j$. Here, $\chi_j$, $\varepsilon_\text{c,j}^\text{max}$, and $N^\text{max}_j$ represent fit parameters for the respective cusp. The given approach is simple and in general describes the data quite well. The cusp-like behavior was also observed in our previous measurements on the compensation of excess micromotion due to dc-stray fields \cite{Harter2013}. Our fit results for the optimal amplitudes for the electric field compensation are $\varepsilon_\text{c,v}^\text{max}=4.2\pm 0.4\:\text{Vm}^{-1}$ and $\varepsilon_\text{c,h}^\text{max}=0.6\pm 0.7\:\text{Vm}^{-1}$, respectively. We note that although at these fields the atomic loss is minimized, it still remains at a level of about $15\%$, which is mainly due to other uncompensated excess micromotion. For comparison, the atom loss in the absence of an ion after a hold time of 1$\:$s is only about $3\%$.

To check for consistency we carry out model calculations for the remaining number of atoms $N$ as a function of micromotion energy. In our model we take into account atom loss due to elastic atom-ion collisions with a loss rate $\Gamma_\text{el}$ [see Eqs.$\:$(\ref{eq:elasticrate}) and (\ref{eq:elastic})]. This elastic rate depends on the average two-body collision energy $E_\text{col}=(1-\mu/m_\text{Rb})\times E^\text{kin,a}$, where $m_\text{Rb}$ is the mass of a $^{87}$Rb atom. Here, $E^\text{kin,a}$ is a function of the electrical field amplitudes $\varepsilon_\text{c,j}$. We calculate $E^\text{kin,a}$ using Eq.$\:$(\ref{eq:muEnergy}) and $E^\text{kin,a}(\varepsilon_\text{c,j}=0)=4\:\text{mK}\times k_\text{B}$. Furthermore, we include background atom loss with a rate of about $\Gamma_\text{bg}=500\:\text{s}^{-1}$ in our model. The rate equation for the atom loss reads
\begin{equation}
\label{eq:rateequation}
\dot{N}=-\frac{\Gamma_\text{el}+\Gamma_\text{bg}}{N_0}N\,,
\end{equation}
where $N_0$ is the initial atom number. This yields the solution
\begin{equation}
\label{eq:remaining}
N(t)=N_0\,\text{exp}\left( -\frac{\Gamma_\text{el}+\Gamma_\text{bg}}{N_0}t\right)\,.
\end{equation}
The red dashed lines in Figs.$\:$\ref{fig:phasemini}a and b show the results of our model calculations using Eq.$\:$(\ref{eq:remaining}). In order to ease the comparison between the theory curve and the experimental data, we scaled the initial atom number by a factor 0.98 for the calculations, which is, however, still well within the uncertainty of our atom number calibration. Figure $\:$\ref{fig:phasemini} shows that after this scaling the agreement between our model calculations and the experimental data is reasonably good.

\subsection{Compensation of rf-induced axial micromotion}
\label{sec: axial}

\begin{figure}[t]
	\begin{center}
		\includegraphics[width=\columnwidth]{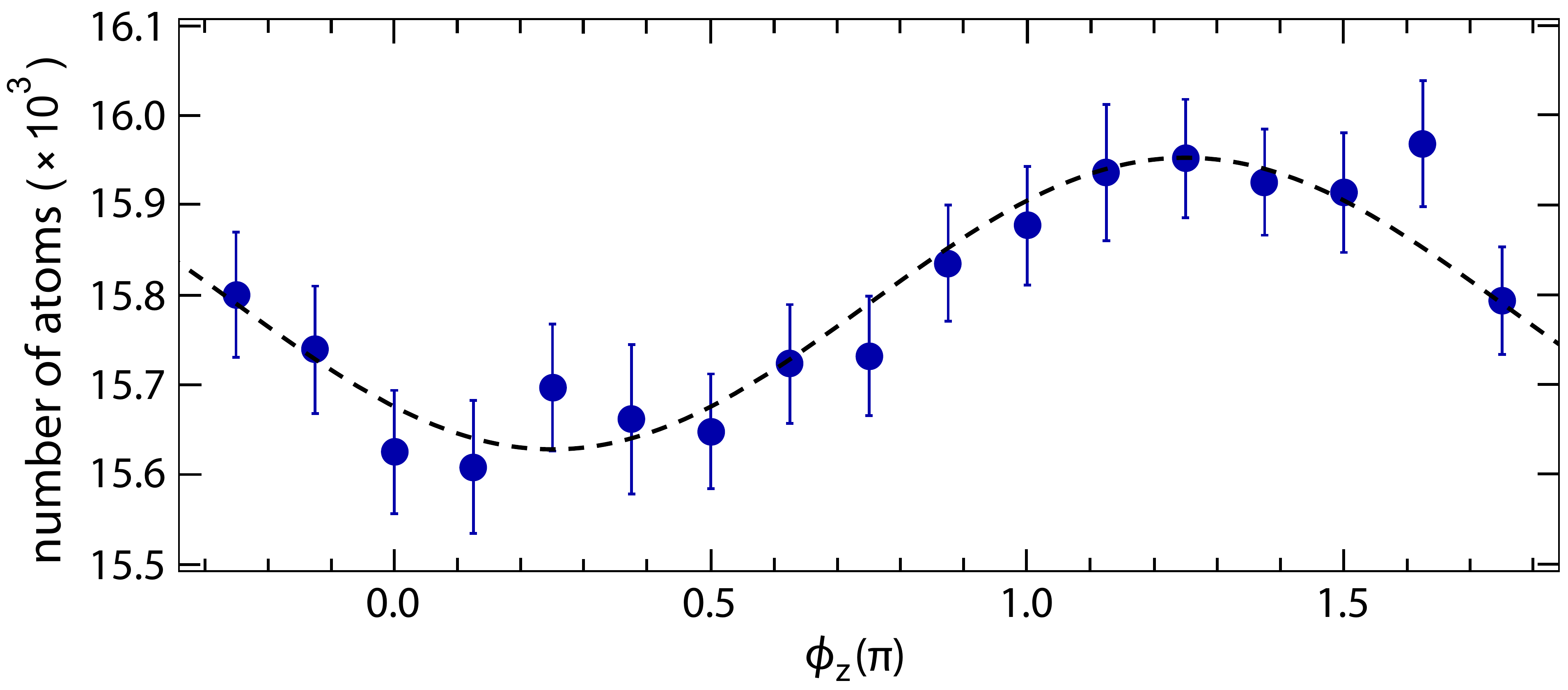}
		\caption{Remaining atom number after 500$\:$ms of interaction time with the ion as a function of the phase $\phi_z$ of the ac-voltage applied on the endcap electrode. Blue dots are the measurements. The error bars indicate the $1\sigma$ statistical uncertainty. Each data point is the average value of $140$ experimental runs. The dashed curve is a fit of a sine function, $\propto \text{sin}(\phi_z + \Delta\phi_z)$, where $\Delta\phi_z = (-0.77\pm 0.02)\pi$.
		}
		\label{fig:axphase}
	\end{center}
\end{figure}

In order to compensate for rf-induced axial excess micromotion, we apply [in accordance with Eq.$\:$(\ref{eq:acmicro})] an ac-voltage of $V_z=V_{0,z} \, \sin(\Omega t+\phi_z)$  to one of the endcap electrodes. We search for the optimal amplitude $V_{0,z}$ and  phase $\phi_z$, again using the scheme of minimizing atomic losses. For this we work with an atom cloud of  $1.7\times 10^4$ atoms and with an ion trap where phase micromotion is not compensated, i.e. $V_\text{c,v}=V_\text{c,h}=0$. We start by optimizing $\phi_z$. The voltage amplitude is set to a fixed value of $V_{0,z}=1\:\text{V}$ which corresponds to an electric field amplitude of about $\varepsilon_{0,z}=8\:\text{Vm}^{-1}$ at the position of the ion in the trap center. Figure \ref{fig:axphase} shows the measured remaining number of atoms as a function of the phase $\phi_z$ for an interaction time of 500$\:\text{ms}$ with the ion. We can fit a sine function, $\propto \text{sin}(\phi_z + \Delta\phi_z)$, to the data and obtain $\Delta\phi_z = (-0.77\pm 0.02)\pi$. The atomic losses are minimal for $\phi_z = -\Delta\phi_z + \pi/2$.

\begin{figure}[t]
	\begin{center}
		\includegraphics[width=\columnwidth]{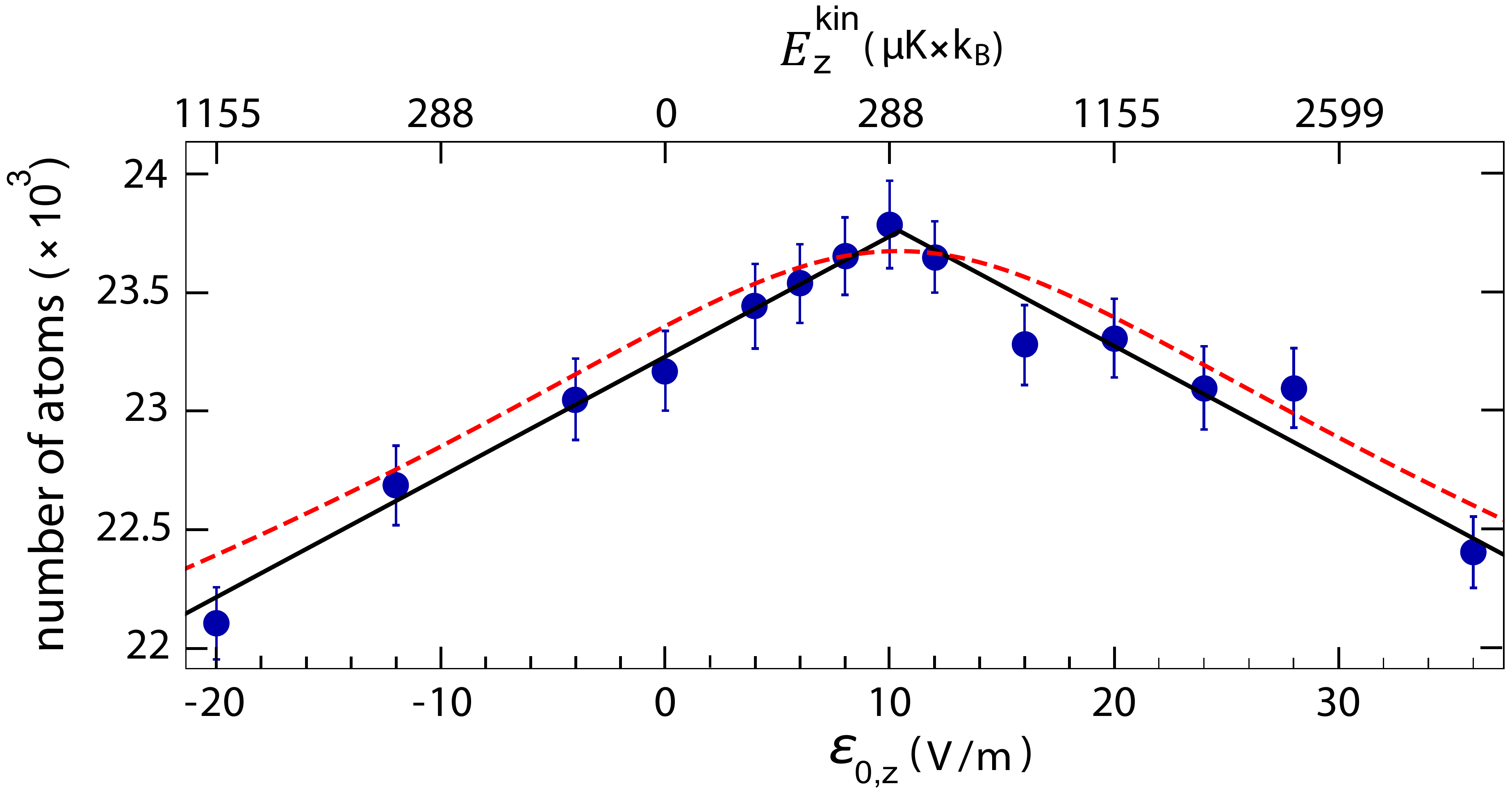}
		\caption{Remaining number of atoms after interaction with the ion as a function of the electric field amplitude $\varepsilon_{0,z}$. This field amplitude is produced at the position of the ion by applying an ac-voltage to one of the endcap electrodes. Blue data points are measurements and the error bars represent the $1\sigma$ statistical uncertainty. The black solid line is a cusp function fit and the red dashed line shows the expected atom number obtained from model calculations.
		}
		\label{fig:axamp}
	\end{center}
\end{figure}

We fix this phase for the search of the optimal voltage amplitude $V_{0,z}$, which is carried out next. This time, we work with a configuration where phase micromotion is already compensated for as discussed in subsection \ref{sec: phase}. Now, initial atom densities of about $n_\text{at}=3.6\times 10^{11}\:\text{cm}^{-3}$ are used. Figure \ref{fig:axamp} shows the remaining atom number for an interaction time of $1\:\text{s}$ as a function of the electric field amplitude $\varepsilon_{0,z} = V_{0,z} \times 8\:\text{m}^{-1}$. As in Fig.$\:$\ref{fig:phasemini},  the top abscissa indicates the electric field amplitude in terms of a corresponding micromotion energy $E^\text{kin}_z$ according to Eq.$\:$(\ref{eq:muEnergy}). To determine the optimum compensation voltage amplitude we fit  the cusp function $-\chi |\varepsilon_{0,z}- \varepsilon_{0,z}^\text{max}|+N^\text{max}$ to the data (see solid lines in Fig.$\:$\ref{fig:axamp}) and obtain $\varepsilon_{0,z}^\text{max}=10.4\pm0.5\:\text{Vm}^{-1}$. This electric field amplitude corresponds to $V_{0,z} = 1.3\pm 0.06\:\text{V}$. Furthermore, the given value for $\varepsilon_{0,z}^\text{max}$ corresponds to a decrease in micromotion energy of almost $300\:\upmu\text{K}\times k_\text{B}$. This is about six times larger than the energy regarding phase micromotion, as discussed in subsection \ref{sec: phase}. Again, the red dashed line in Fig.$\:$\ref{fig:axamp} represents the result of model calculations using Eq.$\:$(\ref{eq:remaining}). Here, we take into account that phase micromotion was already compensated. As before, we applied a 0.98 scale factor to the initial atom number in the calculations. The agreement between the model calculations and the experimental data is again quite good.

\section{Probing micromotion compensation via reactive atom-ion collisions}
\label{sec: inelastic}
\begin{figure}[t]
	\begin{center}
		\includegraphics[width=\columnwidth]{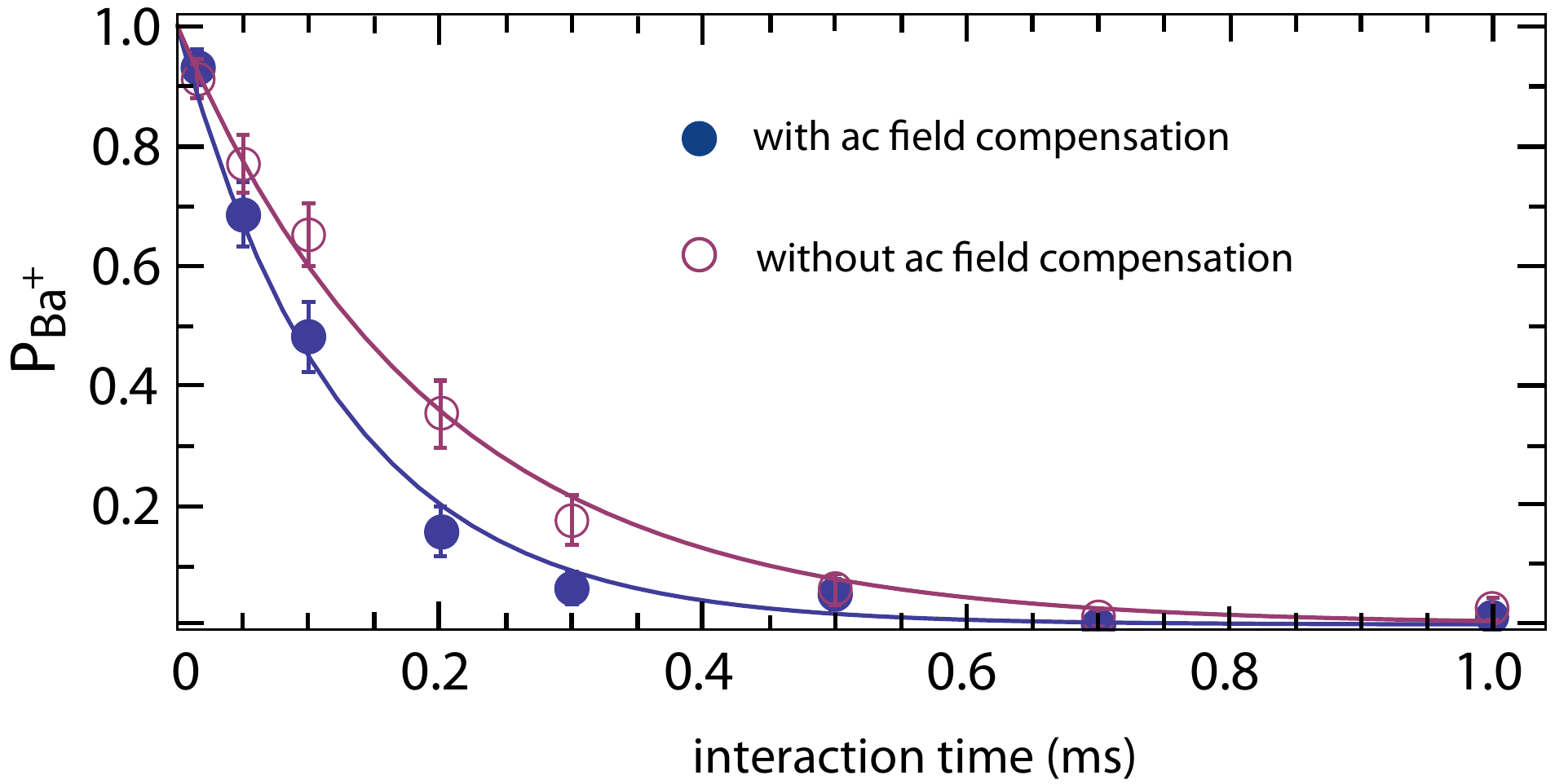}
		\caption{Effect of micromotion compensation for three-body recombination. Shown is the probability $P_{\text{Ba}^+}$ that a Ba$^+$ ion has not reacted away after
			being immersed into an ultracold cloud of atoms for an interaction time $t$.
			The purple empty circles (blue filled circles) are measurements carried out without (with) applying the micromotion compensation. The error bars give the $1\sigma$ statistical uncertainty. The continuous lines are exponential fits.
		}
		\label{fig:ratecomparison}
	\end{center}
\end{figure}

We now probe micromotion via reactive collisions instead of elastic ones. For this, we work with large atomic densities of about $n_\text{at}=7\times 10^{13}\:\text{cm}^{-3}$ where three-body recombination, $\text{Ba}^++\text{Rb}+\text{Rb}\rightarrow (\text{BaRb})^++\text{Rb}$, is by far the dominant reaction process \cite{Krukow2016a}. The three-body recombination rate is given by $k_3 \, n_\text{at}^2$ and the rate constant $k_3$ scales as \cite{Krukow2016a,Krukow2016b}
\begin{equation}
\label{eq:react_rate_const}
k_3\propto \widetilde{E}_\text{col}^{-3/4}\,.
\end{equation}
Here, $\widetilde{E}_\text{col}$ is the three-body collision energy in the center-of-mass frame.
Since the kinetic energies of the atoms can be neglected as compared to the ion energy, the average three-body collision energy is given by
\begin{equation}
\widetilde{E}_\text{col} = \left(1-\frac{m_\text{Ba}}{m_\text{Ba} + 2m_\text{Rb}} \right) E^\text{kin,a} = 0.56E^\text{kin,a}\,.
\end{equation}
As stated earlier at $\widetilde{E}_\text{col}=2.2\:\text{mK} \times k_\text{B}$ the rate constant is $k_3=1.04\times 10^{-24}\:\text{cm}^6\text{s}^{-1}$. Eq.$\:$(\ref{eq:react_rate_const}) shows that the reaction rate will strongly scale with the micromotion energy of the ion. We measure the reaction rate as follows. The ion is immersed into the atomic cloud for a variable time $t$. Afterwards, we use near-resonant fluorescence imaging for a duration of 100$\:$ms to detect the ion. If no cold Ba$^+$ ion is detected, we infer that a reaction has occurred. After repeating the experiment 90 times we obtain a probability that a reaction has taken place within a given time $t$. Figure \ref{fig:ratecomparison} shows the probability $P_{\text{Ba}^+}$ that the Ba$^+$ ion has not reacted as a function of $t$. The open purple circles correspond to a measurement without compensation of phase and of rf-induced axial excess micromotion. In contrast, the filled blue circles represent a measurement where we have compensated micromotion, as described in sections \ref{sec: phase} and \ref{sec: axial}. The inelastic rate clearly increases when micromotion is compensated, as expected.

The decays can be well fit by exponentials $P_{\text{Ba}^+}=\text{exp}(-\Gamma_\text{inel}t)$, where $\Gamma^{\text{w}}_\text{inel}=(8.0\pm 0.6)\times 10^3\:\text{s}^{-1}$ and $\Gamma^{\text{wo}}_\text{inel}=(5.2\pm 0.3)\times 10^3\:\text{s}^{-1}$ for the case with (w) and without (wo) micromotion compensation, respectively. Using Eq.$\:$(\ref{eq:react_rate_const}) we obtain
\begin{equation}
\frac{\widetilde{E}_\text{col}^\text{w}}{\widetilde{E}_\text{col}^\text{wo}}=\left( \frac{\Gamma_\text{inel}^\text{wo}}{\Gamma_\text{inel}^\text{w}}\right)^{4/3} = 0.56 \pm 0.07 \,.
\end{equation}
Therefore, the compensation of phase and rf-induced axial micromotion in our setup reduces
the original kinetic energy $E^\text{kin,a}_\text{wo}$ of the ion by about $0.44E^\text{kin,a}_\text{wo}$.

In sections \ref{sec: phase} and \ref{sec: axial} we have determined that this reduction of excess micromotion energy is $350\:\upmu\text{K}\times k_\text{B}$, which corresponds to a decrease of $\Delta E^\text{kin,a}=5\times 350\:\upmu\text{K}\times k_\text{B}=1.75\:\text{mK}\times k_\text{B}$. Since $0.44E^\text{kin,a}_\text{wo}=1.75\:\text{mK}\times k_\text{B}$, we obtain $E^\text{kin,a}_\text{wo}\approx 4\:\text{mK}\times k_B$. This is in agreement with the value stated in section \ref{sec: phase}. The given comparison demonstrates the good consistency between the two methods.

The remaining kinetic energy $E^\text{kin,a}$ of about $2.2\:\text{mK}\times k_\text{B}$ after compensation in our setup is still substantial. It can probably only partially be explained by collisionally induced micromotion for which the typical energy scale is expected to be on the order of $40\:\upmu\text{K}\times k_\text{B}$ for our trap parameters \cite{Cetina2012}. We are planning to investigate this in detail in the near future.

\section{Summary and Discussion}
\label{sec:summary}

In conclusion, we demonstrate how to compensate rf-excess micromotion by studying elastic or reactive collisions of an ion with ultracold neutral atoms. We minimize unwanted rf-electrical fields down to the level of about $1\:\text{Vm}^{-1}$, which is of similar quality as achieved via other conventional compensation schemes \cite{Chuah2013, Tanaka2012, Narayanan2011, Keller2015, Meir2016}. In our setup the compensation decreased the ionic excess micromotion energy by about $350\:\upmu\text{K}\times k_\text{B}$. Furthermore, we deduce from known scaling laws of collision rates that the ion after full dc- and rf-field compensation still has a substantial amount of kinetic energy of about $2.2\:\text{mK}\times k_\text{B}$ when located inside the cold atomic gas. This residual kinetic energy might be partially explained by collision-induced micromotion. In the future it will be interesting to investigate this fundamental limit in more detail, e.g. by varying trap parameters of the Paul trap such as the $q$ and $a$ parameters. Due to characteristic scaling properties of the relevant energy terms with $q$ and $a$ this will allow for discriminating between different sources of micromotion \cite{Haerter2013}.

The method discussed here is especially convenient for atom-ion hybrid systems since both species are readily available. Compensating excess micromotion allows for reaching low collisional energies between atom and ion. This could be of interest e.g. in the search for shape resonances in atom-ion collisions, see e.g. \cite{Silva2015}. Finally, it has been predicted \cite{Cetina2012} that the $s$-wave collisional regime can be reached for large atom-ion mass ratios, e.g. as for Yb$^++$Li \cite{Fuerst2018}, since here micromotion-induced heating is comparatively small. For further suppressing micromotion-induced heating, Rydberg dressing of atoms \cite{Secker2017} could be applied.

\section*{Acknowledgments}
This work was supported by the German Research Foundation (DFG, Deutsche Forschungsgemeinschaft) within SFB/TRR21 and grant DE 510/2-1.

\section*{Appendix: Some general considerations on compensating excess micromotion}
\label{sec:Appendix}

Here, we discuss in a very general way excess micromotion in a Paul trap\footnote{This trap does not have to exhibit perfect rotational symmetry.}. The motion of a confined ion is determined by the exposure to constant and rf-electrical fields\footnote{We restrict the discussion to the quasi-static regime where dynamical coupling of $\vec{\mathcal{E}}$- and $\vec{\mathcal{B}}$-fields (e.g. induction) can be neglected.}. Concretely, we consider the three electrical fields $\vec{\mathcal{E}}_c, \vec{\mathcal{E}}_{\cos}, \vec{\mathcal{E}}_{\sin}$, characteristic for a Paul trap. $\vec{\mathcal{E}}_c (\vec{r})$ is time-independent, while $\vec{\mathcal{E}}_{\cos}(\vec{r}) \propto \cos(\Omega t)$ and $\vec{\mathcal{E}}_{\sin}(\vec{r}) \propto \sin(\Omega t)$ are quadrature components of the rf-field. The electrical fields $\vec{\mathcal{E}}_c,$ $\vec{\mathcal{E}}_{\cos}, \vec{\mathcal{E}}_{\sin}$ can each be Taylor expanded around the trap center position $\vec{r}_0$. For each expansion the first term is a homogeneous offset field and  the second term is a quadrupole field, followed by higher multipole terms such as the octupole field. In the Paul trap it is the quadrupole field which is used for trapping. For excess micromotion to vanish at $\vec{r}_0$, we need $\vec{\mathcal{E}}_c (\vec{r}_0) = \vec{\mathcal{E}}_{\cos}(\vec{r}_0) =  \vec{\mathcal{E}}_{\sin} (\vec{r}_0) = 0$, which is equivalent
to the vanishing of the respective offset field terms of the Taylor expansions. The remaining quadrupole fields at location $\vec{r}$ in the direct vicinity of $\vec{r}_0$  will generally dominate over the higher multipole fields, as long as $\vec{r}- \vec{r}_0$ is much smaller than the distance to any of the trap electrodes. The total quadrupole field $\vec{\mathcal{E}}_\text{qp}(\vec{r})$ (i.e. the sum of the three quadrupole fields)  is fully determined by the second derivatives of the corresponding electrostatic potential $\phi$, i.e. $\vec{\mathcal{E}}_\text{qp}(\vec{r}) = H(\phi)  (\vec{r}-\vec{r}_0) $. Here, $H(\phi)$ is the Hessian matrix of $\phi$, i.e. $H_{i,j}(\phi)={\partial^2 \phi \over \partial x_i \partial x_j}$, where $x_i, x_j \in \{ x, y, z \}$. Since $H(\phi)$ is symmetric it can be diagonalized. In the corresponding coordinate system $\{x', y', z'\}$ the electrical quadrupole field can be written as $\vec{\mathcal{E}}_\text{qp} = a x' \hat x' + b y' \hat y' + c z' \hat z'$, similarly as in Eq.$\:$(\ref{eq:field1}). Here, $a,b,c$ are time-dependent coefficients. Therefore, the motions of the ion along directions $\hat x', \hat y', \hat z'$ are decoupled and can be described by Mathieu equations. Thus, the main requirements for a Paul trap are fulfilled.

In order to cancel the offset field components of each of the $\vec{\mathcal{E}}_c, \vec{\mathcal{E}}_{\cos}, \vec{\mathcal{E}}_{\sin}$ fields at location $\vec{r}_0$ we can use three  compensation electrodes to which we apply suitable dc- and rf-voltages. These electrodes produce electrical fields at $\vec{r}_0$ which are preferentially (but not necessarily) mutually perpendicular to each other.

\end{document}